# HUMAN MOBILITY MINING THROUGH HEAD/TAIL BREAKS

Karim Keramat Jahromi,

Department of computer science, University of Milan, Via Comelico 39/41, 20135, Milan, Italy
*Karim.Keramat@unimi.it*

**Abstract**

Nowadays as the world population has become more interconnected and is relying on faster transportation methods, simplified connections and shorter commuting times, we witness a rapid increase in human mobility. In this situation unveiling and understanding human mobility patterns have become a crucial issue to support decisions and prediction activities when managing the complexity of the today's social organization. In practice, the mobility pattern of each individual person consists of the sequence of visited places (Points of interest). Those places and their correlations represent the foundation of most modelling and activity researches for understanding human mobility. Even though visited places underpin almost the majority of works in this field, their features remain largely unknown because in previous works, they have been mainly considered as uncharacterized spot points in an area or social gathering places, without considering the roles and importance of places to the behavior of each single user.

In this paper, a framework to deal with the importance of places from perspective of an individual mobile user is proposed. Considering that most of the features and properties of human mobility show heavy tail trend because of the hierarchy structure and high level of heterogeneities in people's movement, therefore, we are motivated to use Head/Tail breaks for mining and classification of mobility features instead of applying the classical approaches such as K-Means, etc, since we expect that the Head /Tail breaks can better capture and retrieve the heterogeneity in people's movement. On the other hand, in recent years the large availability of smartphones combined with the pervasive availability of communication infrastructures, gave rise of collecting the datasets recording human mobility, providing basis to validate our work on larger scale. Pilot datasets that used for analysis are Cellular and Wifi datasets collected from smartphones.

## INTRODUCTION

Understanding the rules that govern human mobility is a crux of many studies in multidisciplinary fields, such as urban planning, traffic forecasting and the spreading of biological and computer viruses (Csáji et al., 2013; Frias-Martinez et al., 2011; González et al., 2008; Scafetta, 2011). Human mobility also determines the formation of social aggregations, so that its awareness is prominent for understanding how specific social networks form and grow (Scafetta, 2011). It has also been a topic of many studies in mobile networks because it is at the heart of the decision of the next hop by predicting the next opportunity in the design of routing protocols for Opportunistic and Delay Tolerant Networks (Karamshuk et al., 2011).

We need to observe how people move in order to capture the mobility patterns and to study the human movement. It is hard and even against of privacy policies to keep continuous track of individuals at all times. In situations that smartphones and also the majority of portable wireless devices are carried by humans, these mobile devices indicate the mobility behavior of the people carrying them. All this has a significant impact on network operation and performance. To some extent, we can, apparently, capture mobility patterns of roaming users in statistical terms. This is relevant in attempting to reveal the mobility patterns related to human behavior, so as to achieve a more realistic mobility model and thereby predict movements of users that can be exploited in different domains and applications (Pirozmand et al., 2014).

On the other hand, the current existence technologies which act as proxies to observe human movements such as Cellular, WiFi networks, and GPS, provide various levels of accuracy and granularity, with different level of spatio-temporal granularity. In this paper by defining a regularity metric for visiting significant places of the individual user, we characterize this regularity feature from the perspective of individual users through applying Head/Tail breaks classification.

RELATED WORK

In contrast to the Gaussian way of thinking that relying on characterizing things by well-defined Mean and Variance, in fact, there are so many things in the world that cannot be characterized just by well-defined Mean which implies the implication of far more small thing than the large one. So authors in the paper (Jiang, 2015a) proposed an implication of Paretian way of thinking that underlines heavy-tailed behavior and skewed distribution for the better understanding of geographic space scale. Heavy-Tailed distributions (Zipf's family) such as power law distribution which is sometimes referred as scale-free distribution implies a lack of average for characterizing the sizes of things. Relying on this concept, authors (Jiang, 2015a, 2015b) proposed a new classification scheme for data with a heavy-tailed distribution called Head/Tail breaks; those above the mean in the head, and those below the mean in the tail. Most of the previous works in the field of mobility analyzing and characterizing, confirm the heavy-tailed behavior of human mobility features (González et al., 2008; Jia et al., 2012; Noulas et al., 2012; Scafetta, 2011). So relying on this heavy-tailed behavior that is the result of heterogeneity and hierarchical structure in human mobility, we have been motivated to use the Head/Tail breaks instead of using classical mining approach such as K-Means clustering (Keramat Jahromi et al., 2016) for characterizing mobility features that indicate a Heavy-tailed trend. So here we are applying the Head/Tail breaks on a relevance metric defined for characterizing the importance of visited places from the perspective of an individual user. Relevance ratio is defined per user and places. It means that in the first step, we classify users in different Groups according to their behaviors in visiting places and then the visited places by an individual user in each group are mined in different classes.

MOBILITY DATASETS AND PREPROCESSING

In this work, we analyzed a widely used WiFi campus-based dataset collected from the log of association/disassociation of Access Points (APs) at Dartmouth university campus (Kotz et al., 2009) for duration three months. Also, two further cellular network mobility datasets (CDR) have been used. When a mobile user makes or receives a phone call, text SMS or accessing the internet, a CDR is recorded. The CDR data has become a powerful tool to analyze human behavior patterns and an increased interest towards making use of CDRs to analyze the human mobility cheaply, frequently and especially at a very large scale has been recorded recently. Our work relies on two cellular network datasets provided by a cellular network operator (Keramat Jahromi et al., 2016). CDR-17 covers 17 consecutive days and recorded the contact activities(phone call, text SMS or accessing the internet) over the whole metropolitan area, i.e. the city of Milan and its district while CDR-67 covers 67 consecutive days and is only limited to the Milan city. These collected datasets capture the movement of a heterogeneous population and cover a large metropolitan area. Moreover, the number of people involved is consistently greater than other datasets, especially in the cellular network datasets. These features make these datasets suitable resources to be investigated for proposing and validation of mobility models in the metropolitan area.

To extract meaningful visited places which are called Points of Interest (PoIs) and infer informative mobility patterns, we preprocessed both CDR and WiFi datasets.

In cellular networks to extract mobility characteristics of individuals, we need to have enough CDR sample records to study the movement of users, so we filtered out users whose their records data were too short. This means that we chose users with at least one contact activity per day to have enough records of mobility. Also, we combine call/SMS and Internet traffic records to get more data about users' positions. This way, we can consider as Points of Interest (or even Regions of Interest) for a user, the cells he/she visits, i.e. where he/she performs an on-the-phone activity. So the PoI in CDR datasets is cell coverage area that its radius changes from few hundred meters to a kilometer.

To extract significant PoIs from the WiFi dataset, we filtered out APs where the user is just passing throw by considering only APs with **Pause Time > 15 min**. Here in contrast to the (Jia et al., 2012), we used stay time duration (Association time to the APs) in the coverage area of APs, instead of a number of visits to APs.

Table 1 indicates the summary about datasets before and after preprocessing.

**Table1:** Summary about datasets

| Datasets | Number of Users | | Number of Days | Number of PoIs |
|---|---|---|---|---|
| | Before preprocessing | After preprocessing | | |
| CDR-17 | 1,291,416 | 543,085 | 17 | 12,898 |
| CDR-67 | 734,149 | 17,400 | 67 | 5,398 |
| WiFi | 17,404 | 14,082 | 90 | 972 |

**DAILY MOBILITY REGULARITY**

We know that people do not move randomly; by contrast, their movement is influenced by their needs, commitments, and their social ties. As a result, mobility patterns show daily regularity and periodicity (Hasan et al., 2012; Hsu et al., 2007). Authors in (Keramat Jahromi et al., 2016) have characterized these regularities in visiting places by defining the Relevance Ratio (RR) of the Point of Interest (PoI) P for a user u as:

$$RR(P, u) = \frac{d_{visit}(P, u)}{d_{total}(u)}$$

Where $d_{visit}(P, u)$ is the number of days when a given place $P$ has been visited by user $u$ and $d_{total}(u)$ is the number of days which is recorded in the user's dataset. The day was adopted as temporal window since it represents the fundamental period when considering life routine of individuals. The relevance ratio (RR) captures the probability of how likely an individual will move towards a place or return back to it according to his/her movement history. As the importance of a place for a user is revealed by the frequency s/he happens to visit it (Jia et al., 2012), we resort to using relevance to measuring it. The empirical relevance ratio CCDF distributions obtained from introduced datasets (after preprocessing) aggregated over all users and visited PoIs is shown in Figure 1.

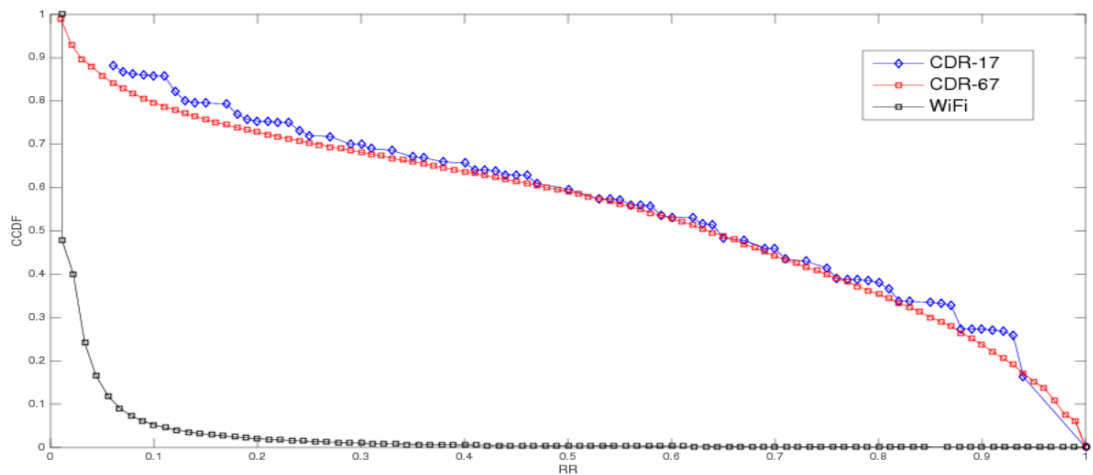

**Fig. 1.** The empirical distributions of RR for three different datasets

As can be observed in Figure 1, in CDR datasets there is a high number of PoIs that are sporadically visited (small value of RR) while a few PoIs are visited almost daily (high value of RR). In the longer duration dataset; the CDR-67, this pattern is highlighted as there was the time to record the high number of PoIs that users visit one or very few times by chance or by the special event and just a few PoIs that visited regularly and frequently. These kinds of observations are remarking the head-and-tail analogy (Jiang, 2015a) that can be widely seen even in other examples such as national telecoms vs. Skype, Encyclopedia Britannica vs. Wikipedia, and also in the human mobility properties. So the heavy-tail behavior of distributions implies **Paretian** thinking that believes in far more small things than large ones (Jiang, 2015a, 2015b). Here we apply Head/Tail breaks on a relevance ratio per each user for clustering visited PoIs by users. The number of times that scaling pattern of far more small things than large one recurs on relevance ratio metric plus one, is referred as *ht-index*, which determines the number of groups that users who have visited PoIs, could be mined. For instance Group N, includes people that their visited PoIs can be classified into N separate classes. Table 2 indicates the percentages of users that classified in each group.

As we can observe from Table 2, the percentage of users in each Group for different datasets is different. In WiFi dataset majority of the population belong to Group 1, means classified just in one category, and this result makes sense considering the mobility of all users in the limited geographical area of the campus environment.

Taking into account that the spatial granularity of PoIs is wider in CDR dataset w.r.t the WiFi dataset, so we observed different Head/Tail classification in WiFi and CDR datasets. In the former case, an urban PoI coincides with a cell tower and approximates a hexagon with a few hundred meters side. While a PoI is extracted from the WiFi dataset, is restricted to AP with radius coverage area of maximum 25 meters. So there is a possibility that several nearby PoIs which have been extracted from WiFi dataset and are in the coverage area of one cell tower in CDR dataset, all to be merged and recognized as one PoI in CDR dataset.

Table 2. Head/Tail Breaks classification.

| Number of Users | 12,267 | 146,326 | 17,400 |
|---|---|---|---|
| Group = ht-index | *WiFi Dataset* | *CDR-17 Dataset* Users (%) | *CDR-67 Dataset* |
| 1 | 91.253 | 14.550 | 3.000 |
| 2 | 1.087 | 12.310 | 62.370 |
| 3 | 5.920 | 69.315 | 33.340 |
| 4 | 1.740 | 3.760 | 1.234 |

In CDR-17 and CDR-67 datasets, most of the population mined in Group 2 and Group 3, respectively. Among the emerged groups by above Head/Tail breaks classification, here we focus on Group 3 in WiFi, and CDR-17 datasets and also Group 2 and 3 in the CDR-67 dataset, since these groups cover the almost majority of the users population in these datasets.

The adoption of a Head/Tail breaks for detecting the relevance classes allows us to adaptively select their bounds and avoid the choice of fixed thresholds. In fact, the application of Head/Tail clustering best suits the diverse human mobility patterns and mitigates the spatio-temporal heterogeneity which characterizes the different datasets. In comparison to the result of mining through K-means clustering approach presented in (Keramat Jahromi et al., 2016), Head/Tail breaks better perform mining and retrieve the hierarchal structure and high level of heterogeneities in people's movement. The K-means clustering approach proposed in (Keramat Jahromi et al., 2016) could just retrieve three groups while the Head/Tail breaks by mining users in four groups in all datasets (Table 1), better reveals the hierarchal in people's movement. For instance, by applying the K-means mining to the WiFi dataset (Keramat Jahromi et al., 2016) only 44.86% of users population could be mined in three groups (since large portion of population are sedentary) while in Head/Tail breaks all population of users were mined.

The distributions of Relevance Ratio (RR) in different classes in Group 3 for WiFi and CDR-17 datasets and also in Group 2 and 3 of the CDR-67 dataset have been indicated in below figures, respectively.

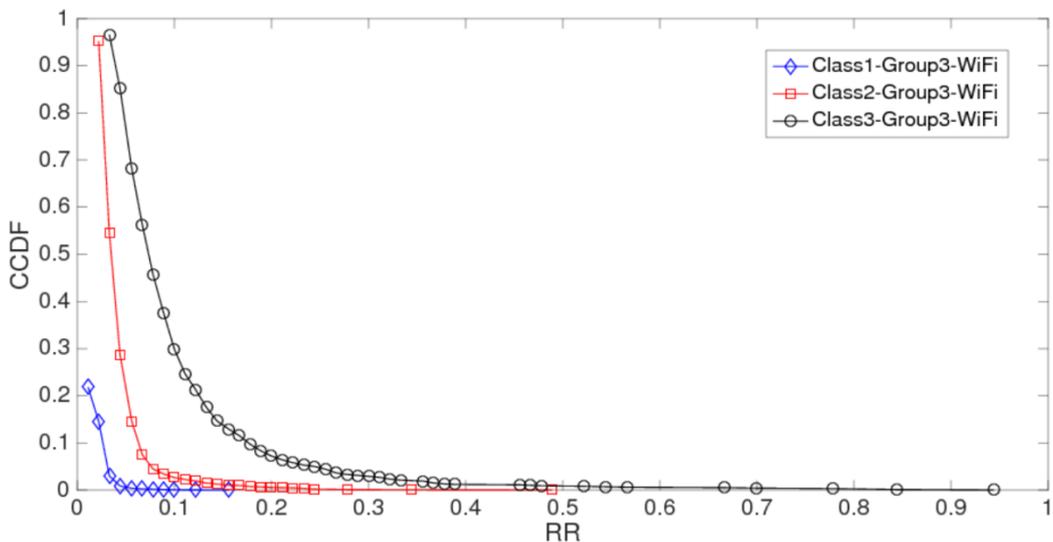

**Fig. 2.** The empirical distributions of RR for different classes in group 3 for the WiFi dataset.

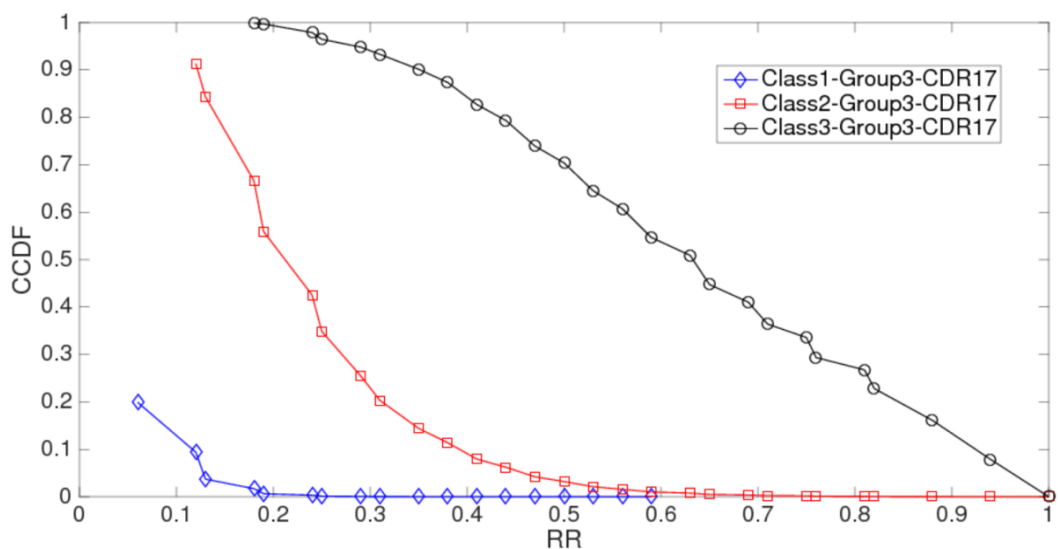

**Fig. 3.** The empirical distributions of RR for different classes in group 3 for the CDR-17 dataset.

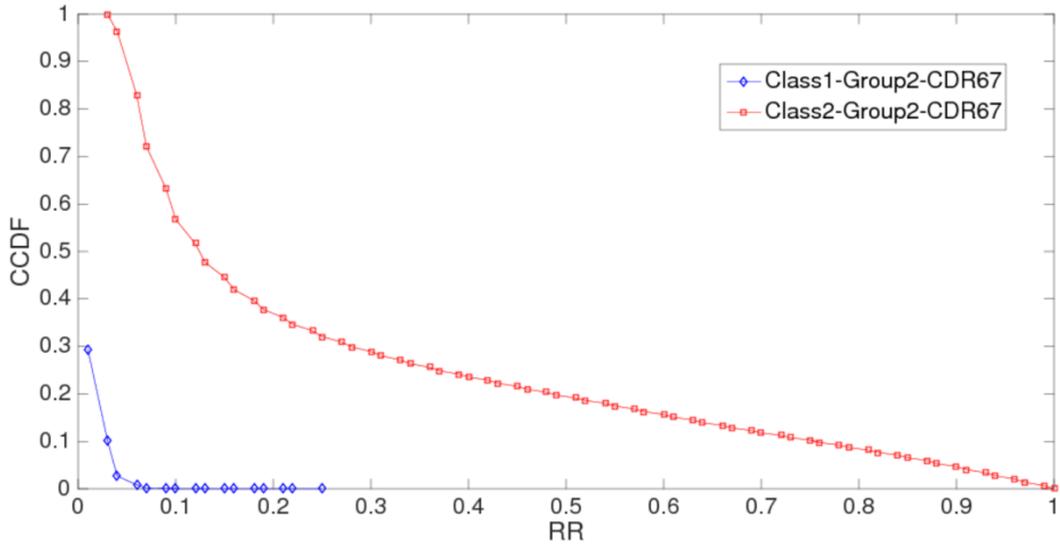

**Fig. 4.** The empirical distributions of RR for different classes in group 2 for the CDR-67 dataset.

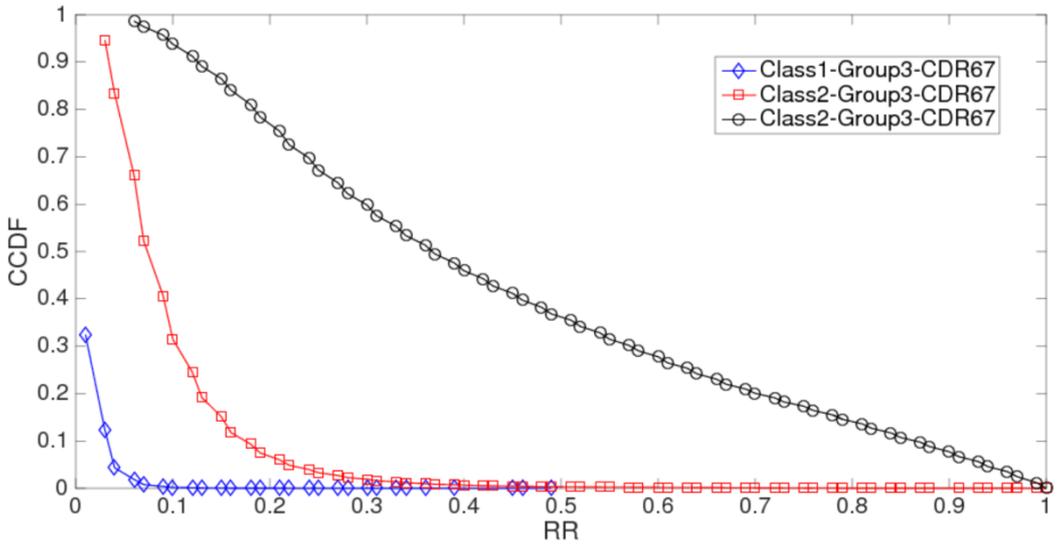

**Fig. 5.** The empirical distributions of RR for different classes in group 3 for the CDR-67 dataset.

In all above figures, the empirical relevance distributions reveal the high level of the separability of relevance classes for each dataset. Besides, in class 3, the relevance is much higher than class 1 and class 2, accounting for PoIs actually visited very frequently and regularity, versus the two other classes which are visited occasionally and exceptionally.

PoIs which belong to the class 3 are the PoIs that a person visits more regularly with high Relevance ratio like, for instance, home and workplaces and are called as ***Most Visited Points (MVPs)***. PoIs which belong to the class 2 are those PoIs that are visited less frequently and are known as ***Occasionally Visited Points (OVPs)***. And finally the PoIs in class 3, which correspond to seldom visited PoIs, ***Exceptionally Visited PoIs (EVPs)***. Once we extracted the relevance classes, we focus on the number of distinct visted PoIs in each relevance class.

In Figures 6-9, we show the number of distinct visited PoIs per user (aggregated over all users) associated to each class of relevance in WiFi, CDR17 and CDR 67 datasets, respectively. For all datasets, we observe a remarkable difference between the number of PoIs in EVP class and the PoIs into the other relevance classes (OVP, MVP). This implies the general user's habit to visit many new locations, but also that they regularly move towards very few of them. If we focus on OVP and MVP classes, it turns out that the number of visited OVPs and MVPs are more limited.

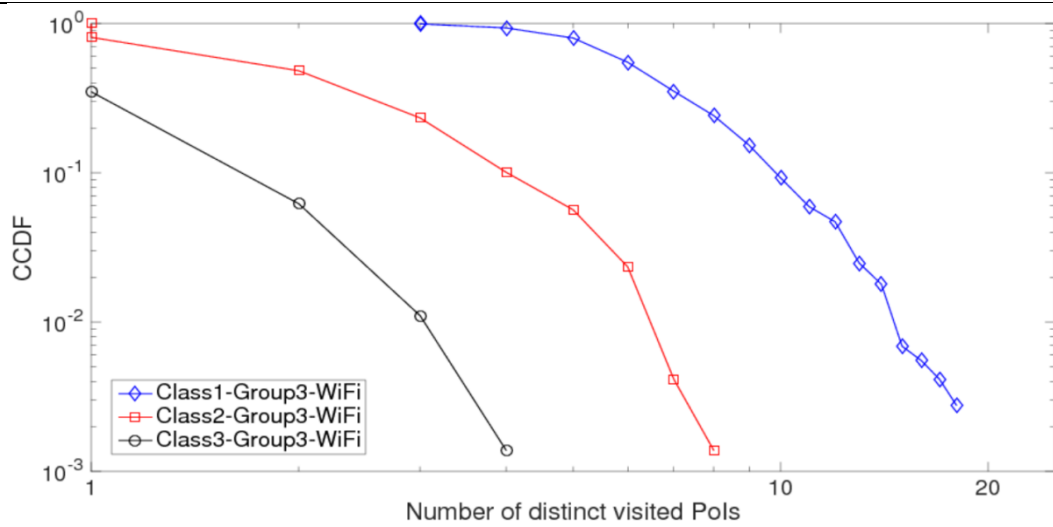

**Fig. 6.** The empirical distributions of the number of distinct visited PoIs per users in different classes in group 3 for the WiFi dataset.

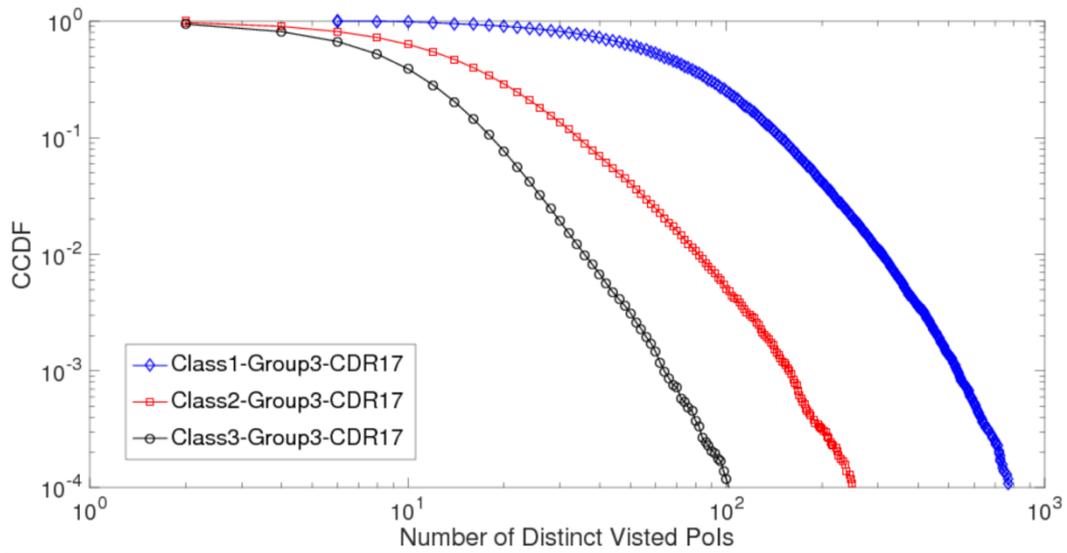

**Fig. 7.** The empirical distributions of a number of distinct visited PoIs in different classes in group 3 for the CDR-17 dataset.

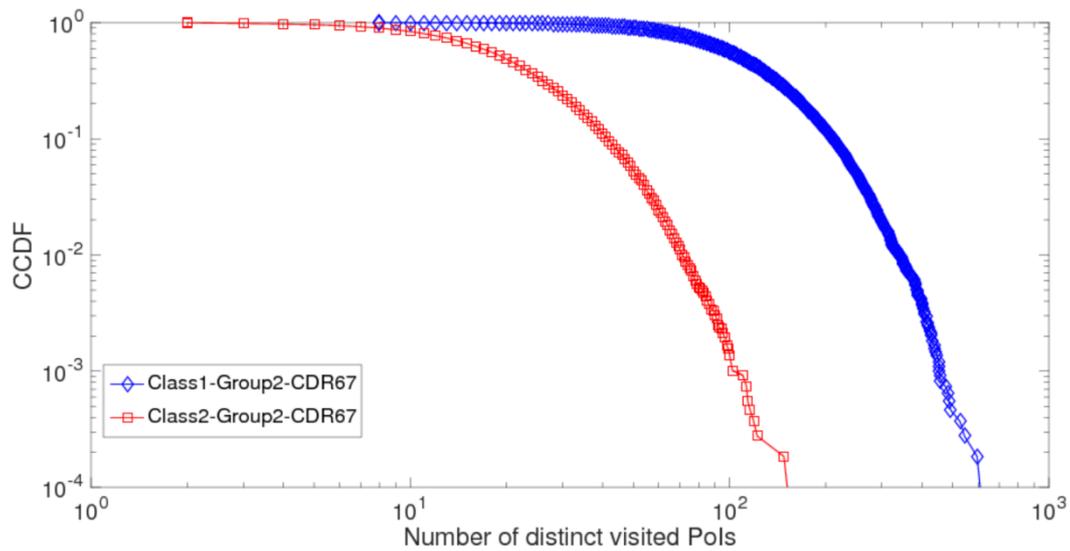

**Fig. 8.** The Empirical distributions of a number of distinct visited PoIs per user in different classes in group 2 for the CDR-67 dataset.

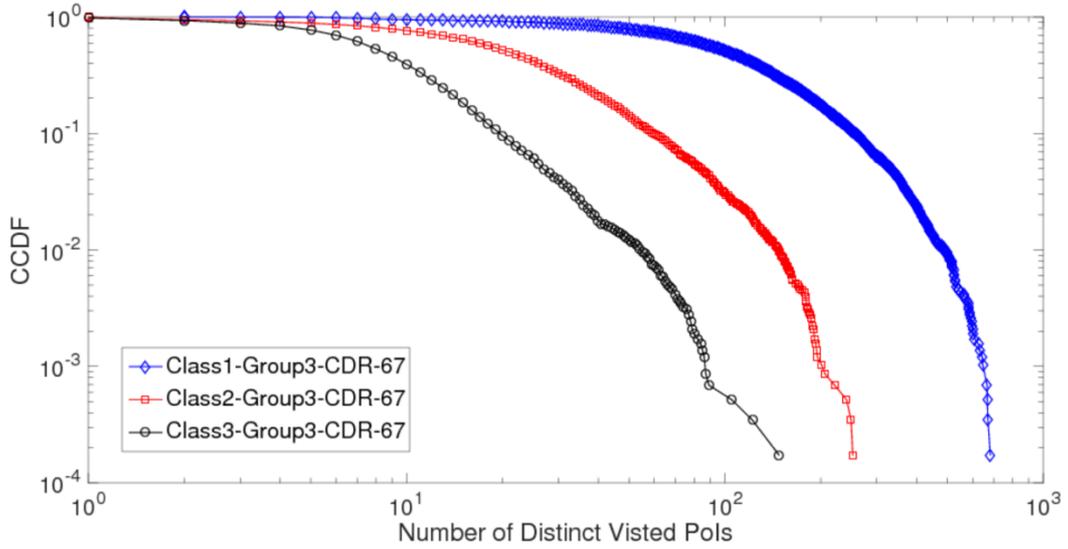

**Fig. 9.** The empirical distributions of a number of distinct visited PoIs per user in different classes in group 3 for the CDR-67 dataset.

The above analysis emphasies the fact that each user has a very small number of favorite places (MVP) which are visited daily (e.g., home, workplace), and a higher, but still limited number of PoIs (OVPs) which are visited with lower frequency (e.g., gym, favorite pub, parent's house). The observed characteristic in MVP class, with a heavy tail distribution of a number of visited PoIs, implies that the majority of users visit just a few places more frequently and regularly which is well aligned with location preference property in human mobility (González et al., 2008; Hsu et al., 2007).

Finally, we enhance the generalizability of the feature of relevance class throughout different datasets by analyzing the average of the percentage of PoIs in the three classes, as reported in Table 3. The behavior is quite similar for all datasets. Most PoIs belong to the EVP class; there are very few MVPs, while OVPs account for a number of places similar to the MVPs class.

**Table 3:** The average percentage of PoIs in each class relevance in group 3

| Dataset | $P_{EVP}\%$ | $P_{OVP}\%$ | $P_{MVP}\%$ |
|---|---|---|---|
| **WiFi** | 64.4 | 23.6 | 12 |
| **CDR-17** | 72.0 | 17.0 | 11 |
| **CDR-67** | 75.6 | 16.5 | 9 |

We can, therefore, conclude that the PoIs classification we identified in terms of relevance by Head/Tail breaking approaches (MVPs, OVPs, EVPs), is generally significant since the distributions of the number of distinct visited PoIs per-user associated to each class of relevance is similar, across datasets with very different characteristics. We have shown that, independently of the dataset characteristics, the places visited by people fall mainly in the EVP class. However, most of the people spend most of their time in MVPs or OVPs; many of them can be found more than half of the time in MVPs. Figure 10 depicts the empirical pause time distributions in different classes of group 3 for WiFi dataset. We can observe a positive correlation between the relevance ratio and pause time duration, that means Pause time duration in MVP class is longer than OVP class, and in OVP is longer than EVP.

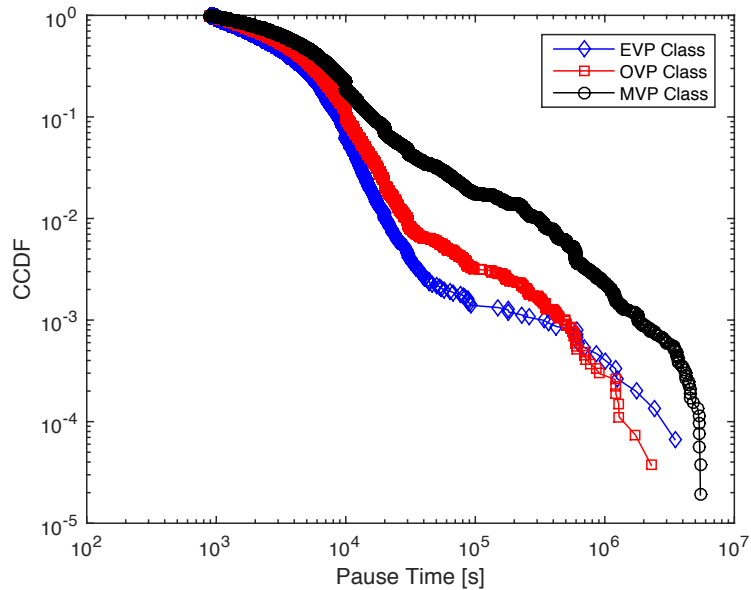

**Fig. 10.** The empirical pause time duration distributions of PoIs in different classes of group 3 for WiFi dataset.

## CONCLUSION

As a result of this work, some interesting properties about human mobility emerge. In fact, it turns out that people visit many places in their life, but they have a very small number of preferred places (MVPs) which are visited daily (e.g., home, workplace), and a higher, but still limited, number of places of interest (OVPs) which are visited with a lower frequency (e.g., gym, favorite restaurant, parent's house). MVPs are PoIs that people spend most of their time. This indicates that those PoIs are the ones that better represent and characterize our life.

Also, this is a proof that Head/Tail breaks can classify very well the heterogeneity and hierarchy wrapped in human mobility.

This work supports its findings by validating results on three datasets with different characteristics in terms of spatial and temporal distribution of the visited places (comparing to the classical K_means clustering approach presented in the (Keramat Jahromi et al., 2016)). According to the extracted results of Head/Tail breaking approach, we can demonstrate the independence of our results with respect to datasets and a specific setting, and we are able to extract a deeper understanding of human mobility.

These novel results can change how mobility is analyzed and modelled. Indeed, we argue that to produce more realistic mobility traces, a mobility model needs to consider the new PoIs classifications introduced herein, and their different classes, their relationships and even transition laws among them. As for relevance to Social Networks and Travel behavior, our results could impact on several areas as:

Characterizing the single individual's mobility and human mobility modelling (Keramat Jahromi et al., 2016), as mobility can be described in terms of regular movement among MVPs, OVPs and EVPs; localization (Zhang et al., 2015) where it can be predicted the probability that people are in MVPs; social interaction studies and data offloading (Rebecchi et al., 2015), as people tend to meet more frequently people with some MVPs in common.